\documentstyle[12pt,epsf,amsfonts,pslatex]{article}
\begin{document}
 \title{Amplification by stochastic interference}
\author{K. Svozil\\
 {\small Institut f\"ur Theoretische Physik}  \\
  {\small Technische Universit\"at Wien   }     \\
  {\small Wiedner Hauptstra\ss e 8-10/136}    \\
  {\small A-1040 Vienna, Austria   }            \\
  {\small e-mail: svozil@tph.tuwien.ac.at}
\\
{\small and}
\\
D. Felix \\
{\small Department of Neurobiology,
University of Berne}\\
{\small Erlachstra\ss e 9a,
CH-3012 Berne, Switzerland}\\
{\small and}
\\
K. Ehrenberger \\
{\small ENT Department,
University of Vienna}\\
{\small W\"ahringer G\"urtel 18-20,
A-1090 Vienna, Austria}}
\maketitle
\newpage

\begin{abstract}
{\em  A new method is introduced to obtain a strong signal by the
interference of weak signals in noisy channels.
The method is based on the
interference of $1/f$ noise from parallel channels.
One realization of
stochastic
interference
is the auditory nervous system.
Stochastic
interference
may have broad potential applications in the information transmission
by parallel noisy channels.}
\end{abstract}

The method of stochastic interference
has been conceived originally for the information processing in the
auditory nervous system \cite{svoz-ehr}.
It makes use of the random fractal geometry of the
spike discharge patterns \cite{liebovitch:88,king:91,teich:89,teich:90a}
which are
processed
by diverging and converging information networks of the auditory system.
This method is distinct from stochastic resonance \cite{stoc-res}, but
when both methods are combined, a fascinating new model of
transsynaptic information transfer emerges \cite{ehrenb-svozil}.

Here, we are interested in more general aspects of stochastic
interference. The method can be sketched as follows. Consider an
information transmission via multiple channels.
Assume further that the information is coded in statistically
self-similar, random
\cite{chaitin-66c,chaitin3,kolmogorov1,schnorr1,martin-lof,ml:70,ml:71}
fractal, patterns \cite{svozil-93}.
The idea that information is coded in the dimensional geometry of random
fractals is not entirely new
\cite{liebovitch:88,king:91,teich:89,teich:90a}.
But here, $n$ fractal information signals (with the same dimensional
parameter) are combined by logical
``and''-operations (equivalent to the set theoretic intersection) to
form a new signal. The new signal has also a fractal geometry. Its
fractal dimension varies $n$ times as strong as variations of the
dimensional parameter of the primary signal.
Thus, when multiple
information channels are combined properly, arbitrary
weak variations of their input signals can be amplified to arbitrary
strong variations of the resulting output channel.

Stochastic interference
operates with $1/f^\beta$ noise \cite{dutta:81,ziel:79},
characterized by
a power spectral density of
$S_V(f)\propto 1/f^\beta$.
This noise corresponds to
a signal $X(t)$ at time $t$ whose graph $\{ (t,X(t))\mid t_{\rm min}\le
t\le t_{\rm max}\}$ has a random fractal geometry. The
fractal (box-counting)
dimension of the graph can be approximated by
\cite{voss85,falconer2}
\begin{equation}
D =\min \{ 2,
 E+{3-\beta \over 2}\} \qquad ,
\end{equation}
where $E$
is the (integer) dimension of the noise.
For onedimensional noise, $E=1$.
White noise corresponds to
$\beta =0$,
brown noise corresponds to
$\beta =2$,
whereas systems showing  $1/f$-noise operate at approximately $\beta
=0.8-1.2$.

Consider a sequence of zeros and ones which constitutes a fractal
pattern. Such a
random fractal of dimension $D$ can, for instance, be
recursively generated by starting with a sequence of ones.
Then, the sequence is subdivided into $k$ blocks of sequences of length
$\delta$ symbols. Then, a fraction of
$1-\exp [(D-1)\log (k)]$
blocks of length $\delta$ symbols is filled with zeros (instead of
ones). Now take the remaining pieces of the pattern containing ones and
repeat the same procedure again (the length of the blocks
decreases by a factor of $k$, until one arrives at $\delta=1$
\cite{falconer2}.

The fractal
dimension of a random fractal signal can be understood as follows.
Divide a sequence of zeros and ones again into $k$ blocks of length
$\delta$. Count how many of these blocks contain ones at all (or, more
realistically for practical applications, up to a density $s$). If $r$
is the number of filled blocks, then
the fractal (box counting) dimension is  given
by
\begin{equation}
D= {\log r\over \log (1/\delta )},
\end{equation}
independent of the scale resolution $\delta$. The fractal dimensional
measure
$D$ should be robust with respect to variations of methods to determine
it. That it, it should remain the same, no matter by which method it is
inferred.

Information can be coded by the random fractal patterns of $1/f$ noise;
in particular by variations of the dimension parameter.
More precisely,
assume, for example, two source symbols $s_1$ and $s_2$
encoded by ($RFP$ stands for ``random
fractal pattern'')
\begin{equation}
\# ( s_i) =\left\{
 \begin{array}{l}
RFP \quad {\rm with}\quad  0\le D(RFP)<D_c\quad {\rm if }\quad s_i= s_1
\\
RFP \quad {\rm with}\quad  D_c\le D(RFP)\le E \quad {\rm if }\quad s_i=
s_2
\\
\end{array}
 \right.
\quad ,
\end{equation}
where $D_c$ is a ``critical
dimension parameter.''

As has been pointed out by K. J. Falconer
\cite{falconer2},
under certain ``mild side conditions,'' the intersection of two random
fractals
$A_1$ and
$A_2$ which can be minimally embedded in ${\Bbb R}^E$ is again
a random fractal with dimension
\begin{equation}
D(A_1\cap A_2)=  \max \{0,D(A_1)+D(A_2)-E \} \quad .
\label{l:a1a2}
\end{equation}
By induction, Eq. (\ref{l:a1a2}) generalizes to the intersection of
an arbitrary number of random fractal sets. Thus,
 the dimension of the intersection
 of $n$ random fractals ${\cal A}=\{A_1,\ldots ,A_n\}$ is
 given by
\begin{equation}
D({\cal A})=D(\; \bigcap_{i=1}^n A_i\;)= \max \{0,
-E(n-1) + \sum_{i=1}^nD(A_i)\}\quad
.
\label{mrfp}
\end{equation}

We shall concentrate on the case of onedimensional signals where $E=1$.
Assume that the signals are represented by sequences of zeros and ones.
Assume further
 $n$ random fractal signals $A_1,\ldots ,A_n$.
Each one of these
sequences is transmitted in a separate channel.
The sequences are then
recombined to form a new, secondary signal sequence.
In particular, we shall be interested in the {\em intersection} of
$n$ signals encoded by random fractal patterns.
An intersection between two signals
$A_1=a_{11}a_{12}a_{13}
\ldots a_{1m}$ and
$A_2=a_{21}a_{22}a_{23}
\ldots a_{2m}$ of length $m$, $a_{ij}\in \{0,1\}$,
is again a signal
$A_1\cap A_2 =A_3=a_{31}a_{32}a_{33}
\ldots a_{3m}$
of length $m$ which is defined by
\begin{equation}
a_{3i} =\left\{
 \begin{array}{l}
1 \quad {\rm if}\quad  a_{1i}a_{2i}=1 \quad {\rm and}
\\
0 \quad {\rm otherwise.}
\\
\end{array}
 \right.
\label{l-5}
\end{equation}
We shall denote this setup by the term {\em stochastic
interference.} Taking the product in (\ref{l-5}) amounts to the logical
``and'' operation, if 0 and 1 are identified with the logical values
``false'' and ``true,'' respectively.

Let us discuss shortly two features of {\em stochastic interference.}
Firstly,
the combination of white noise, denoted by
${\Bbb I}$ with $D({\Bbb I})= 1$, with a random fractal signal $A$
results in the recovery of the
original fractal signal with the original dimension; i.e.,
Eq. (\ref{mrfp})
 reduces to
\begin{equation}
D(A \cap {\Bbb I})= D(A)+D({\Bbb I})-1= D(A)\quad .
\end{equation}
Stated pointedly: besides a reduction of intensity, white noise does not
affect the coding.

\begin{figure}
\begin{center}
$\,$
  \epsfysize=8 true cm
  \epsffile{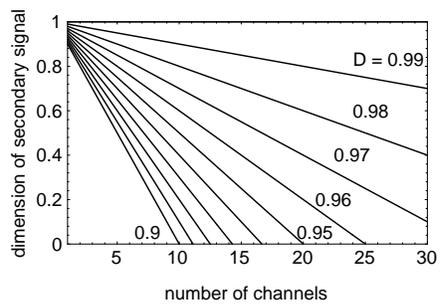}
\end{center}
\caption{Theoretical prediction of the
dimension of the secondary signal
$D({\cal A}) $ as a function of the number of channels $n$
for various values of the
dimension of the primary signal $D$.
\label{ch-mc}}
\end{figure}

Secondly,
by assuming that all $n$ random fractals have equal dimensions, i.e.,
$D(A_i)=D$ for $1\le i \le n$,
 Eq. (\ref{mrfp})
 reduces to
\begin{equation}
D({\cal A})= \max \{ 0,n(D-1)+1\} \quad .
\label{mrfp1}
\end{equation}
In Fig.
\ref{ch-mc}, $D({\cal A})$ is drawn for various dimensions $D$ as a
function of the number of channels $n$.
An immediate consequence of Eq. (\ref{mrfp1}) is that, for truely
fractal signals ($D< 1$), any variation of the fractal dimension of the
secondary signal
is directly proportional to the number $n$ of
the primary
signals; i.e.,
\begin{equation}
\Delta D({\cal A}) = n\Delta D\quad {\rm for}\quad  D\neq 1\quad .
\end{equation}
Therefore, the more channels there are, the more the
dimension of the secondary source
varies in response to variations of the primary source;
there is an ``amplification'' of any change in the primary signal.

This amplification, however, has a price:
any increase
in the amplification of the variation of the primary
dimension obtained by additional channels results in a reduction
of the overall secondary signal strength.

\begin{figure}
\begin{center}
$\,$
  \epsfysize=8 true cm
  \epsffile{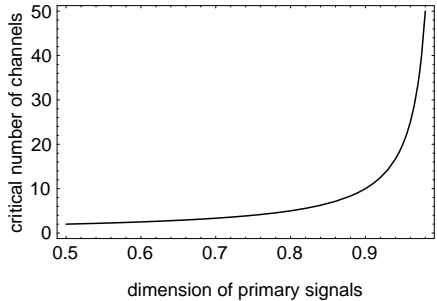}
\end{center}
\caption{   Theoretical prediction of
the critical number of channels as a function of the
dimension of the primary signal.
\label{ch-mulcr}}
\end{figure}

In Fig. \ref{ch-mulcr}, the number of critical channels, for which the
secondary signal vanishes (all $a_{3i}=0$), is drawn against the
dimension of the primary signals.
One arrives at the number of critical channels $n_c$ by setting
$D({\cal A})=0$ in
Eqn. (\ref{mrfp1}) and solving for $n$. That is,
\begin{equation}
n_c ={1\over 1 -D} \mbox{ for }0\le D< 1.
\end{equation}
For a channel number of
$10-20$, the fractal
dimension of the primary signal has to be within the $0.9-1$-range
in order to balance the attenuation.

We close this short discussion of stochasic interference by pointing out
the possibility of a twofold information transfer in one and the same
system of multiple noisy
channels: firstly, transfer by the standard coding
techniques \cite{hamming}, and secondly, modulated by it, transfer by
information coding using
1/f noise with stochastic interference.
This form of double-band information transfer may be realized in the
auditory pathway of mammals and has potential applications in
communication technology as well.


\end{document}